# Relativistic motions

A. LOINGER

Dipartimento di Fisica, Università di Milano

Via Celoria, 16 − 20133 Milano, Italy

**Abstract.** − A physical consequence of a well-known Fermi's theorem: no motion of masses can generate gravitational waves.

**1**. − Fermi's geometrical theorem [1] as generalized by Eisenhart [2] affirms: For a manifold endowed with a *symmetric* connection it is possible to choose a coordinate system with respect to which the components $\Gamma^i_{jk} (= \Gamma^i_{kj})$ of the connection are *zero* at *all* points of a curve (or of a portion of it).

For a Riemann-Einstein spacetime this means that there exists a coordinate system $(z)$ with respect to which the first derivatives of the components $h_{jk}(z), (j,k = 0,1,2,3)$, of the metric tensor are *zero* at all points of a curve (or of a portion of it), in particular at all points of a time-like world line.

**2**. − Let us now consider a *continuous medium* (for instance, a perfect fluid) characterized by a certain mass tensor $T_{jk}$, $(j,k = 0,1,2,3)$, and let $g_{jk}(x)$ be the solutions of Einstein equations

$$(2.1) \qquad R_{jk} - \frac{1}{2} g_{jk} R = -\kappa T_{jk}$$

corresponding to a generic motion of our medium with respect to a given reference system $(x) \equiv (x^0, x^1, x^2, x^3)$. Let us suppose to follow the motion of a given *mass element* describing a certain world line *L*. If we refer this motion − from the initial time $t_0$ on − to Fermi's coordinates $(z) \equiv (z^0, z^1, z^2, z^3)$, the components $h_{jk}(z)$ of the metric tensor will be equal to some constants for *all* points of line *L*. This means that the gravitational field *on L* has been obliterated. Consequently, no gravitational wave has been sent forth. Now, line *L* is quite generic, and therefore *no motion of the continuous medium can give origin to a gravitational radiation.* −

The absence of a "mechanism" apt to generate gravitational waves can be proved also by other arguments ([3], [4], [5], [6]). In paper [5] I have emphasized, in particular, that any particle of an incoherent "cloud of dust", characterized by a mass tensor

$$(2.2) \qquad T^{jk} = \rho \frac{\mathrm{d}x^j}{\mathrm{d}s} \frac{\mathrm{d}x^k}{\mathrm{d}s}, \quad (j,k = 0,1,2,3),$$

(where ρ is the invariant mass density), describes a *geodesic* line: accordingly, it cannot emit gravitational waves. A simple application: the gravitational motions of the members of solar system.

In paper [6] I have given a very simple argument according to which no motion (gravitational or *non*-gravitational) of a mass point can produce gravitational waves. (The restriction of all the above arguments to the motions of *point* masses is clearly inessential from the conceptual point of view). Of course, the above results destroy the current conviction of the real existence of the gravitational waves [7].

"E sarà mia colpa se così è?"
Machiavelli

**REFERENCES**

[1]  FERMI E., *Rend. Acc. Lincei,* **31**[1] (1922) 21 and 51; LEVI-CIVITA T., *Lezioni di calcolo differenziale assoluto* (Stock, Roma) 1925, p.190; IDEM, *Math. Ann.,* **97** (1926) 291.
[2]  EISENHART L.P., *Non-Riemannian Geometry* (Am. Math. Soc., New York) 1927, p.64.
[3]  SCHEIDEGGER A.E., *Revs. Mod. Phys.*, **25** (1953) 451. The author shows that all the computed radiation terms can be destroyed by suitable coordinate transformations at all stages of his approximation procedure.
[4]  INFELD L. AND PLEBANSKI J., *Motion and relativity* (Pergamon Press, Oxford, *etc.*) 1960, pp. 200 and 201. The authors show that the gravitational radiation can be annihilated by an appropriate choice of the reference system at any stage of their approximation method.
[5]  LOINGER A., *Nuovo Cimento* B, **115** (2000) 679.
[6]  LOINGER A., http://xxx.lanl.gov/abs/physics/0106052 (June 17th, 2001) − a more adequate classification would be gr-qc or astro-ph.
[7]  See, e.g., SCHUTZ B.F., *Class. Quantum Grav.*, **16** (1999), A131.